\def \be {\begin{equation}}
\def \ee {\end{equation}}
\def \bea {\begin{eqnarray}}
\def \eea {\end{eqnarray}}
\def \nn {\nonumber}

\def \rr {\raise.35ex\hbox{\small $\prime$}\kern-.17em{\mbox{\large $\imath$}}}
\def \del {\partial}
\def \dels {\partial\kern-.5em / \kern.5em}
\def \As {{A\kern-.5em / \kern.5em}}
\def \Ds {D\kern-.7em / \kern.5em}

\def \a {\alpha}
\def \b {\beta}

\def \eps {\epsilon}

\def \lam {\lambda}
\def \Lam {\Lambda}

\def \ks {k\hspace{-.47em}/\hspace{.1em}}

\documentclass[10pt]{article}

\usepackage{amsfonts}

\begin{document}
\begin{titlepage}

\begin{center}
\hfill hep-th/0410248\\
\vskip .5in

\textbf{\large
UV-finite scalar field theory with unitarity
}

\vskip .5in
{\large Pei-Ming Ho, Yi-Ya Tian}
\vskip 15pt

{\small Department of Physics,
National Taiwan University,
Taipei, Taiwan, R.O.C.}\\
{\small National Center for Theoretical Sciences at Taipei,
Taiwan, R.O.C.}

\vskip .2in
\sffamily{
pmho@phys.ntu.edu.tw\\
yytian@phys.ntu.edu.tw}

\vspace{60pt}
\end{center}
\begin{abstract}

In this paper we show how to define the UV completion of
a scalar field theory such that it is both UV-finite and perturbatively unitary.
In the UV completed theory,
the propagator is an infinite sum of ordinary propagators.
To eliminate the UV divergences,
we choose the coefficients and masses in the propagator
to satisfy certain algebraic relations,
and define the infinite sums involved in Feynman diagram calculation
by analytic continuation.
Unitarity can be proved relatively easily by Cutkosky's rules.
The theory is equivalent to infinitely many particles
with specific masses and interactions.
We take the $\phi^4$ theory as an example
and demonstrate our idea through explicit Feynman diagram computation.

\end{abstract}
\end{titlepage}
\setcounter{footnote}{0}

\section{Introduction}

Quantum field theories in 4 or higher dimensions generically suffer UV divergences,
and need to be regularized.
As Feynman pointed out in his seminal paper \cite{Feynman}
on Quantum Electrodynamics (QED),
regularization can be carried out by modifying the propagators
with higher derivatives.
For a scalar field, for instance,
one can replace the propagator by
\be \label{Ck}
\frac{1}{k^2 - m^2} \rightarrow f(k^2)\equiv \frac{C(k^2)}{k^2 - m^2},
\ee
where
\be
C(k^2) \rightarrow 0 \quad \mbox{as} \quad k^2 \rightarrow \infty.
\ee
If $C(k^2)$ approaches to zero sufficiently fast at large momenta,
the theory becomes UV finite.
The modification of the propagator can be easily realized by
a higher derivative modification
of the kinetic term in the Lagrangian density.
However, such modifications usually violate unitarity.
It is extremely hard to find a consistent co-existence of higher derivatives and unitarity.
A rare example is the noncommutative Lee model given in \cite{ChongSun}.
It is a non-relativistic model where unitarity is proved to the one-loop order.
(The unitarity beyond one-loop is unclear.)

Unitarity is a consistency condition for any quantum theory.
Formally, a quantum theory is unitary as long as its Hamiltonian is Hermitian
and the Hilbert space has a positive definite norm.
The unitarity we care about in this paper is the perturbative unitarity,
which is a stronger, more technical requirement.
It demands that unitarity is observed order by order
(or even diagram by diagram) in the perturbation theory.
We need perturbative unitarity for the perturbation theory to make sense.

In this paper we will construct propagators
such that the corresponding higher derivative scalar field theory
(we consider $\phi^4$ explicitly)
is both UV-finite and unitary.
Its perturbation theory is equivalent to a theory
with infinitely many scalar fields.
This feature distinguishes our approach from other approaches in the literature.
In \cite{Efimov},
a regularization method was discovered so that
perturbative unitarity can be preserved in the end of renormalization
(after taking the cutoff energy to infinity)
for a large class of nonlocal quantum field theories.
In a more closely related work \cite{Woodard},
a clever method was invented to replace the ordinary propagator
by a new propagator with infinite derivatives but without new poles.
The new nonlocal theory preserves both global and gauge symmetries
of the original theory as well as unitarity.
Comparatively, our approach has the advantage of admitting
simpler interactions.

This paper is organized as follows.
We will first comment on the propagator,
about why it is hard for UV-finiteness and unitarity to compromise (Sec. \ref{prop}).
Next we consider in some detail the $\phi^4$ theory with
a propagator of the Pauli-Villars type (Sec. \ref{phi4}).
Then we explain how to choose the propagator and
use analytic continuation to define
Feynman diagrams such that divergences are avoided and unitarity is preserved
(Sec. \ref{ac}).
Finally we make some remarks in the last section.

Our models are directly defined in the path integral formulation.
We will not carry out the canonical quantization or give the operator formulation
in this paper.

\section{Something about the propagator} \label{prop}

According to the Liouville theorem,
the only bounded analytic functions are constants.
Hence the function $C(k^2)$ in (\ref{Ck})
must diverge somewhere on the complex plane of $k^0$.
We will assume that the divergences are simple poles for simplicity.
\footnote{
There is not much discussion about propagating modes due to
double poles or higher order poles in the literature.
We would like to reserve this possibility for future work.
}
Each pole $k^2 - m^2$ corresponds to a propagating mode
with an on-shell condition
\be
k^0 = E(k_i) \equiv \sqrt{k_i^2 + m^2}.
\ee
In the neighborhood of a pole, the propagator is approximated by
\be
f(k^2) \simeq \frac{c(m^2)}{k^2 - m^2}, \quad \mbox{where} \quad
c^{-1}(m^2) \equiv \left. \frac{d f^{-1}}{dk^2}\right|_{k^2 = m^2}.
\ee
The sign of the coefficient $c(m^2)$ is crucial,
as it reflects the sign of the inner product on the Hilbert space.
If we take the convention that $c(m^2) > 0$ for positive definite norms,
$c(m^2) < 0$ implies a ghost.
Since the first derivative of a continuous function $f^{-1}$ changes sign
from one side of a zero to the other side,
$f^{-1}$ can not be continuous if there is no ghost.
This, for instance, excludes the possibility of
\be
C(k^2) = \frac{1}{k^2 - M^2},
\ee
which was used in \cite{Feynman}.
For $C(k^2)$ of this form,
one can check explicitly that the propagator includes a ghost
\be
f(k^2) = \frac{1}{(k^2 - m^2)(k^2 - M^2)} =
\frac{1}{M^2 - m^2}\left( \frac{1}{k^2 - m^2} - \frac{1}{k^2 - M^2} \right).
\ee
This propagator differs from the prescription of ordinary Pauli-Villars regularization
by merely an overall constant.
Unitarity is violated for energies beyond the ghost mass $M$.
In order to avoid the ghost,
a zero should exist between neighboring poles of $f$
(so that $f^{-1}$ has a pole between neighboring zeros).
We can implant the zero in $f$ by multiplying the propagator by
$(k^2 - \bar{m}^2)$,
with $\bar{m} \in (m, M)$.
Unfortunately, this also restores UV-divergence at the same time,
as the new propagator $f$ approaches to zero no faster than $1/k^2$.
This is the dilemma of field theories:
unitarity and UV-finiteness do not like each other.
The same discussion can be extended to higher derivative
modifications of interactions
and we arrive at the same conclusion.

The starting point of our approach is the same as
Pauli-Villars regularization, that is,
we replace a scalar field's propagator by a superposition of ordinary propagators
\be \label{f}
f(k^2) = \sum_n \frac{c_n}{k^2 - m_n^2}.
\ee
The benefit of this choice is that
we can use Cutkosky's rules \cite{Cutkosky}
to prove perturbative unitarity for generic Feynman diagrams.
Otherwise it is almost impossible to prove unitarity to all orders.
To avoid ghosts, we need
\be \label{cn>0}
c_n > 0 \quad \forall n.
\ee

For a scalar field with only dimensionless parameters,
the naive order of divergence equals its energy dimension.
For a diagram with $L$ loops and $I$ internal lines,
the naive order of divergence is $D = 4L - 2I$.
Since this is also the dimension of the diagram,
the divergent terms are proportional to
\be
\Lam^D, \quad \sum_n c_n m_n^2 \Lam^{D-2}, \quad \cdots, \quad
\sum_n c_n m_n^{D-2} \Lam^2, \quad \sum_n c_n m_n^D \log(\Lam^2).
\ee
All UV-divergences all elimiated if
\be \label{cnmn}
\sum_n c_n m_n^{2r} = 0, \quad \mbox{for} \quad r = 0, 1, \cdots, D/2.
\ee

Another way to see the meaning of (\ref{cnmn}) is to
do the high energy expansion of the propagator
\be
f(k^2) = \sum_n \frac{c_n}{k^2 + m_n^2} = \sum_{r=0}^{\infty}(-1)^r \frac{\a_r}{k^{2(r+1)}},
\ee
where
\be
\a_r \equiv \sum_n c_n m_n^{2r}.
\ee
If (\ref{cnmn}) holds for a larger $D$,
the propagator goes to zero faster when $k \rightarrow \infty$.

In general, if we know the most serious divergence of the original theory,
we can impose (\ref{cnmn}) for a sufficiently large $D$ to remove all UV divergences.
Although (\ref{cn>0}) and (\ref{cnmn}) seem contradictory,
we will consider infinite series and play the trick of analytic continuation
such that both conditions are satisfied.

\section{$\phi^4$ theory} \label{phi4}

In this paper, we take $\phi^4$ theory in $3+1$ dimensions as an example.
It is straightforward to generalize our approach to other scalar field theories.
We leave the study of fermionic fields and vector fields for future works.

The Lagrangian density of the $\phi^4$ theory is
\be
{\cal L} = \phi f^{-1}(-\del^2) \phi + \lam \phi^4,
\ee
where $f$ is defined by (\ref{f}).

\subsection{UV-divergence} \label{UVdiv}

In 4 dimensions the coupling constant $\lam$ is dimensionless,
and so the dimension of a diagram can only come from
the momentum cutoff $\Lam$ and masses $m_n$.
Hence the divergent terms of dimension 2 in a Feynman diagram must be of the form
\be
\label{quad}
a\left(\sum_n c_n\right)\Lam^2 + b\left(\sum_n c_n m_n^2 \log(\Lam^2/m_n^2)\right)
+\mbox{finite},
\ee
where $a$ and $b$ are dimensionless finite coefficients,
possibly including parameters such as $\lam$ and $(\sum_m c_m)$.
Divergent terms of dimension 0 are of the form
\be
\label{log}
\sum_n c_n \log(\Lam^2/m_n^2).
\ee

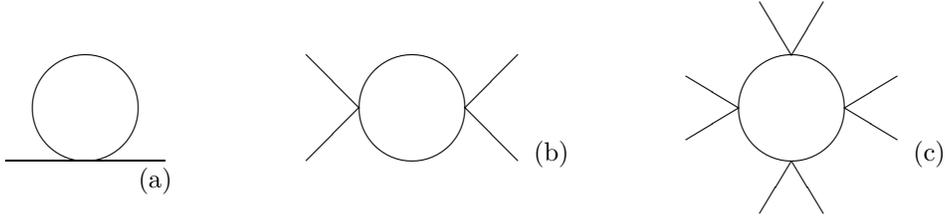
\begin{figure}
\label{fig1}
\setlength{\unitlength}{2pt}
\begin{center}
\begin{picture}(20,30)(-10,-15)
\put(0,0){\circle{20}}
\put(-15,-10){\line(1,0){30}}
\put(10,-15){(a)}
\end{picture}
\hspace{2cm}
\begin{picture}(40,30)(-20,-15)
\put(0,0){\circle{20}}
\put(10,0){\line(1,1){10}}
\put(10,0){\line(1,-1){10}}
\put(-10,0){\line(-1,1){10}}
\put(-10,0){\line(-1,-1){10}}
\put(23,-10){(b)}
\end{picture}
\hspace{2cm}
\begin{picture}(40,30)(-20,-15)
\put(0,0){\circle{20}}
\put(10,0){\line(5,3){10}}
\put(10,0){\line(5,-3){10}}
\put(0,10){\line(3,5){6}}
\put(0,10){\line(-3,5){6}}
\put(-10,0){\line(-5,3){10}}
\put(-10,0){\line(-5,-3){10}}
\put(0,-10){\line(3,-5){6}}
\put(0,-10){\line(-3,-5){6}}
\put(23,-10){(c)}
\end{picture}
\end{center}
\caption{Some one-loop diagrams}
\end{figure}

Let us compute a few Feynman diagrams explicitly.
First, consider the one-loop diagram in Fig. 1(a).
This diagram has quadratic divergence and
after Wick rotation it is given by
\be
\int d^4 k f(k^2) = \sum_n c_n \int \frac{d^4 k}{k^2 + m_n^2}
= \pi^2 \sum_n c_n \left(
\Lam^2 - m_n^2 \log(\Lam^2/m_n^2) + \cdots \right),
\ee
where terms of order ${\cal O}(\Lam^{-2})$ are omitted,
and $\Lam$ is the cutoff of the Euclidean 4-momentum.
The integral over $k$ is an infinite constant.
To avoid this divergence we want
\bea
&\sum_n c_n m_n^2 = 0, \label{cm} \\
&\sum_n c_n = 0, \label{c}
\eea
If both conditions are satisfied so that the divergent terms can be droped
and we can ignore $1/\Lam^2$ terms by taking the limit $\Lam\rightarrow \infty$,
the diagram becomes
\be
\int d^4 k f(k^2) = \pi^2 \sum_n c_n m_n^2 \log m_n^2.
\ee

The same diagram can be computed with a different regularization scheme.
We can first integrate out $k_0$ using the residue theorem,
and then impose a 3-momentum cutoff $\Lam_3$.
What one obtains this way is
\be
\int d^4 k f(k^2) = \pi^2 \sum_n c_n \left( 2\Lam_3^2 - m_n^2 \log(\Lam_3^2/m_n^2)
- 2 m_n^2 \log 2 + \cdots \right).
\ee
Remarkably, the results of the two regularization schemes are exactly the same
if the same conditions (\ref{cm}) and (\ref{c}) above are satisfied.

Similarly, dimensional regularization gives
\be
\int d^4 k f(k^2) = \pi^2 \sum_n c_n \left(
-2 m_n^2 \frac{1}{\eps} + m_n^2 \log m_n^2 + \cdots \right),
\ee
which is again precisely the same as the other two methods of regularization
if (\ref{cm}) and (\ref{c}) holds.

Let us now consider another one-loop diagram Fig. 1(b).
This diagram is given by
\be
{\cal M} = \int d^4 p d^4 q f(p^2) f(q^2) \delta^{(4)}(p+q-k),
\ee
where $k$ is the total incoming 4-momentum of external legs on the left
(or equivalently the outgoing momentum on the right).
Using Feynman's parameter, we rewrite it as
\be
{\cal M} = \sum_{m,n} c_m c_n \int_0^1 d\a \int d^4 p
\frac{1}{(p^2 - \Delta_{mn}(\a))^2},
\ee
where
\be
\Delta_{mn}(\a) = \a(1-\a) k^2 - \a m_m^2 - (1-\a) m_n^2.
\ee
The integral over $p$ has a logrithmic divergence
\be \label{1lb}
{\cal M} \propto \sum_{m,n} c_m c_n \int_0^1 d\a \log(\Lam^2/\Delta_{mn}),
\ee
where $\Lam$ is the Euclidean 4-momentum cut-off.
Apparently, the divergence can be ignored if (\ref{c}) holds.
The finite piece independent of $\Lam$ is roughly
\be
\sum_{m,n} c_m c_n \left(
\frac{m_m^2 \log(m_m^2) - m_n^2 \log(m_n^2)}{m_m^2 - m_n^2} - 1
+ {\cal O}(k^2/m_n^2) \right),
\ee
where we assumed that $k$ is small compared with the masses for simplicity.

\subsection{Perturbative unitarity}

The $\log$ function becomes complex when $k^2$ gets larger
than the critical energy for pair creation of the lightest particle.
Since our propagator is simply the superposition of ordinary propagators,
Cutkosky's rules apply to the imaginary part of ${\cal M}$
\footnote{
This means that we can replace each term $(k^2 - m_n^2)^{-1}$
by a delta function $\delta(k^2 - m_n^2)$ (up tp $2\pi$).
}
\be \label{imM}
\mbox{Im}{\cal M} \propto \sum_{m,n} c_m c_n \int d^4 p \int d^4 q
\delta_+(p^2 - m_m^2) \delta_+(q^2 - m_n^2) \delta^{(4)}(p + q - k).
\ee
Perturbative unitarity then requires that this is equal to
\be \label{RHS}
\int \frac{d^4 p}{(2\pi)^4} \int \frac{d^4 q}{(2\pi)^4} (2\pi) \delta_+(f(p^2)) (2\pi) \delta_+(f(q^2))
(2\pi)^4 \delta^{(4)}(p + q - k).
\ee
Since
\be
\delta_+(f(k^2)) = \sum_n c_n \delta_+(k^2 - m_n^2),
\ee
unitarity is verified for this diagram.

Actually, the rules of drawing Feynman diagrams
for our higher derivative $\phi^4$ theory is the same as another theory
with the Lagrangian
\be
{\cal L} = \sum_n \frac{1}{c_n}\phi_n\left(-\del^2-m_n^2\right)\phi_n
+ \lam \left(\sum_n \phi_n\right)^4.
\ee
Or equivalently, after a change of variable $\phi_n = \sqrt{c_n} \varphi_n$,
the Lagrangian becomes
\be \label{Ln}
{\cal L} = \sum_n \varphi_n\left(-\del^2-m_n^2\right)\varphi_n
+ \lam \left(\sum_n \sqrt{c_n} \varphi_n\right)^4.
\ee
This theory is obviously unitary to all orders
at least when there are no infinite sums.

\subsection{1-loop diagrams}

The expression for a generic one-loop diagram
(such as Fig. 1(b)(c)) with $V$ vertices is proportional to
\be \label{k20}
{\cal M} \propto
\sum_{n_1 \cdots n_V} c_{n_1} \cdots c_{n_V} \left(
\frac{m_{n_1}^2 \log(m_{n_1}^2)}{\prod_{j = 2}^V (m_{n_1}^2 - m_{n_j}^2)}
+ \mbox{permutations} + {\cal O}(k^2/m_n^2) \right)
\ee
after integrating out the loop momentum and Feynman parameters,
where again we assumed a small $k^2$ expansion for simplicity.
Higher order terms in the small $k$ expansion can also be explicitly computed.
The coefficient of the $(k^2)^m$ term is of the form
\be \label{k2m}
\sum_{n_1 \cdots n_V} c_{n_1} \cdots c_{n_V} \sum_{i=1}^V \left(
\sum_{(r_1 \cdots r_V)}
\frac{C_{i;r_1\cdots r_V}m_{n_1}^{2r_1}\cdots m_{n_V}^{2r_V}}
{\left( \prod_{j\neq i}(m_{n_i}^2-m_{n_j}^2 )\right)^{2m}} +
\sum_{(s_1\cdots s_V)}
\frac{D_{i;s_1\cdots s_V}m_{n_1}^{2s_1}\cdots m_{n_V}^{2s_V}(m_{n_i}^2\log m_{n_i}^2)}
{\left( \prod_{j\neq i}(m_{n_i}^2-m_{n_j}^2 )\right)^{2m+1}}
\right),
\ee
where $C_{i|r_1\cdots r_V}$ and $D_{i|r_1\cdots r_V}$ are numerical constants
and $r_1 \cdots r_V, s_1, \cdots s_V$ are integers satisfying
\bea
&r_i \geq 1 \quad \forall i, \quad \mbox{and} \quad r_1 + \cdots + r_V = m(2V-3)-V+2, \\
&s_i \geq 1 \quad \forall i, \quad \mbox{and} \quad s_1 + \cdots + s_V = m(2V-3).
\eea

\subsection{Generic Feynman diagrams}

A generic Feynman diagram with $L$ loops is of the form
\be
{\cal M} = \sum_{n_1 \cdots n_I} c_{n_1} \cdots c_{n_I}
\int d^4 p_1 \cdots \int d^4 p_L \prod_{i=1}^I \frac{1}{q_i^2 + m^2_{n_i}},
\ee
where $I$ denotes the number of internal lines
and $q_i$ the momentum of the $i$-th internal line,
which is a linear combination of the loop momenta $p_j$
and the momenta of external lines $k_i$.
Using Feynman's parameters, this can be written as
\be
{\cal M} \propto \sum_{n_1 \cdots n_I} c_{n_1} \cdots c_{n_I}
\int_0^1 d\a_1 \cdots \int_0^1 d\a_I \delta(\a_1 + \cdots + \a_I)
\int d^4 p_1 \cdots \int d^4 p_L \frac{1}{\left(\sum_{i=1}^I \a_i(q_i^2+m_i^2) \right)^I}.
\ee
Its dimension is
\be
D = 4L - 2I,
\ee
which can be $2, 0, -2, -4, -6, \cdots$.
By shifting the loop momenta $p_j \rightarrow p'_j$,
the integrand can be simplified as
\be
\frac{1}{\left( \sum_{j=1}^L \b_j {p'_j}^2 + \Delta \right)^I},
\ee
where $\b_j$'s are functions of the parameters $\a_i$, and
\be \label{Delta}
\Delta
= \Delta_0 + \sum_{i, j = 1}^E A_{ij} k_i k_j, \quad
\Delta_0 = \sum_{i = 1}^I \a_i m_{n_i}^2.
\ee
Here $k_i$'s ($i=1,\cdots,E$) denote the momenta of external lines,
and $A_{ij}$'s are functions of the Feynman parameters $\a_i$.
Then for $D < 0$ we can scale $p'_j$ such that
\be \label{general}
{\cal M} \propto \sum_{n_1 \cdots n_I} c_{n_1} \cdots c_{n_I}
\int_0^1 d\a_1 \cdots \int_0^1 d\a_I \delta(\a_1 + \cdots + \a_I)
h(\a_i) \Delta^{D/2}.
\ee
where
\be
h(\a_i) = \int d^4 p_1 \cdots \int d^4 p_L \frac{1}{\left( \sum_{j=1}^L \b_j {p'_j}^2 + 1 \right)^I}.
\ee
$h(\a_i)$ is a regular function of the parameters $\a_i$.

\section{Regularization by analytic continuation} \label{ac}

In the previous section we saw that in order to
avoid the UV divergences,
both conditions (\ref{cm}) and (\ref{c}) must be satisfied.
On the other hand we need $c_n > 0$ for unitarity.
This seems impossible.

We propose to solve this dilemma in the following way.
We shall choose the propagator (\ref{f}) to be an infinite series
of ordinary propagators.
In particular $c_n$'s are certain carefully chosen functions of a parameter $z$,
while $m_n$'s can be fixed.
The infinite series involved in a (regulated) Feynman diagram
should be well defined when $z$ lies within a certain range.
Then we can analytically continuate $z$ to a special value $z_0$ at which
conditions (\ref{cm}) and (\ref{c}) are both satisfied,
and furthermore we need all unregulated terms remain finite.

\subsection{The propagator}

We illustrate this idea with an example.
Let the propagator be defined by (\ref{f}) with
\bea
c_0 &=& \frac{1}{1-e^{-z}}, \\
c_n &=& e^{zn} \quad \mbox{for} \quad n \geq 1, \label{cnexp} \\
m^2_0 &=& \frac{1-e^{-z}}{1-e^{-(z+a)}}, \\
m_n^2 &=& e^{a n} \quad \mbox{for} \quad n \geq 1, \label{mnexp}
\eea
where $a > 0$ is a constant.
The masses are monotonically ordered after we set $z = z_0>0$
\be
m_0^2 < m_1^2 < m_2^2 < \cdots < m_n^2 < m_{n+1}^2 < \cdots.
\ee
When $z < - a < 0$,
both infinite sums in (\ref{cm}) and (\ref{c}) are convergent
and can be easily computed
\be
\sum_{n=0}^{\infty} c_n = c_0 + \frac{e^z}{1 - e^z}, \quad
\sum_{n=0}^{\infty} c_n m_n^2 = c_0 m_0^2 + \frac{e^{z+a}}{1 - e^{z+a}}.
\ee
Note that both expressions are now well defined for any value of $z$
except $z = 0$ and $z = - a$,
so we can analytically continuate $z$ to a positive value $z_0$.
Then both conditions (\ref{cm}) and (\ref{c}) hold,
and we can take the limit $\Lam \rightarrow \infty$ without UV-divergence.
Furthermore, we still have $c_n > 0$ and $m_n^2 > 0$ for all $n$ with $z_0 > 0$.

\subsection{Analytic continuation to $z > 0$}

Although we have now successfully avoided UV-divergences,
we also need all UV-finite terms to remain finite after analytic continuation.
In our model,
each Feynman diagram ${\cal M}$ equals an infinite sum
of the same Feynman diagram in the ordinary $\phi^4$ theory
with different masses.
For a diagram with dimension $D < 0$,
its value decreases when the mass of $\phi$ increases,
thus there is no divergence in the limit $n \rightarrow \infty$
if $m_n$ goes to infinity sufficiently fast
(as in our case).
For these diagrams the infinite sums are convergent when $z < 0$.

Schematically,
diagrams with dimension $D \neq 0$ are of the form (\ref{general})
\be
\sum_{n_1 \cdots n_I} c_{n_1} \cdots c_{n_I} \Delta^{D/2},
\label{D>0}
\ee
where $\Delta$ is given in (\ref{Delta}),
and we have ignored the integration of Feynman parameters.
As $\Delta \sim m_n^2 \sim e^{a n}$ grows exponentially with $n$,
$c_n$ decays exponentially with $n$ for $z < 0$.
For $z$ sufficiently small
\be \label{zsmall}
z < -Da/2
\ee
the series converge.
(For $D = 0$ the factor $\Delta^{D/2}$ in (\ref{D>0})
should be replaced by $\log(\Delta)$,
which grows like $n$ when $n$ increases.
So the series converges for any $z < 0$.)
We conclude that the infinite sums in all Feynman diagrams
are always well defined for $z$ sufficiently small (negative).

The next step is to analytically continuate $z$ to a positive contant $z_0 > 0$.
This may lead to new infinities even for diagrams which
are originally free of UV-divergences.
Let us compute several Feynman diagrams explicitly
to find the condition of convergence.

\subsubsection{One-loop diagrams}

Fig. 1 (a) was computed in Sec. \ref{UVdiv} to be proportional to
\be
\sum_{n = 0}^{\infty} c_n m_n^2 \log m_n^2.
\ee
According to our choice of the coefficients $c_n$ and masses $m_n$
given in (\ref{cnexp}) and (\ref{mnexp}),
this quantity is a convergent series for $z < -a$ which can be easily summed up as
\be
c_0 m_0^2 \log m_0^2 + \frac{a}{4\sinh^2((z+a)/2)}.
\ee
This expression is well defined for any $z \neq -a$,
so we are allowed to analytically continuate $z$ to $z_0$.
We have $z_0 \neq -a$ since both $z_0$ and $a$ are chosen to be positive numbers.

Now we consider Fig. 1(b).
The leading order in the low energy (small $k^2$) expansion is
\bea
{\cal M} &=& \sum_{m, n = 0}^{\infty} c_m c_n \int_0^1 d\a \log(\a m_m^2 + (1-\a) m_n^2) \nn \\
&=& c_0^2 \log m_0^2 + 2 c_0 \sum_{n=1}^{\infty} c_n \left( \log m_n^2 +
\frac{m_0^2 (\log m_n^2 - \log m_0^2)}{m_n^2 - m_0^2} \right) +
\sum_{m, n = 1}^{\infty} c_m c_n \frac{m_m^2 \log m_m^2 - m_n^2 \log m_n^2}{m_m^2 - m_n^2} \nn \\
&=& c_0^2 \log m_0^2 + \frac{a c_0}{2\sinh^2(z/2)} +
\sum_{r=0}^{\infty} m_0^{2r} \left( \frac{2 a c_0 m_0^2}{4\sinh^2((a(r+1)-z)/2)} - \frac{\log m_0^2}{e^{a(r+1)-z} - 1} \right)
+ {\cal M}_1,
\label{MM1}
\eea
where ${\cal M}_1$ will be computed below.
In the above we used the expansion
\be \label{m>n}
\frac{1}{m_n^2 - m_0^2} = \sum_{r=0}^{\infty} m_n^{-2(r+1)} m_0^{2r}
\ee
for $n > 0$,
and then we interchanged the order of the sum over $n$ and the sum over $r$,
which is legitimate because the infinite series converges (for $z < 0$).
After summing over $n$ we arrive at an infinite sum over $r$,
which is a convergent series for generic $z$.
The final expression allows us to analytically continuate $z$ to $z_0 > 0$.
The only restriction is that
\be \label{z0-cond}
z_0 \neq (r+1) a \quad \mbox{for} \quad r = 0, 1, 2, \cdots ,
\ee
otherwise the 3rd term in (\ref{MM1}) diverges.

Similarly we can deal with ${\cal M}_1$, which is
\be
{\cal M}_1 = \sum_{m, n = 1}^{\infty} c_m c_n \frac{m_m^2 \log m_m^2 - m_n^2 \log m_n^2}{m_m^2 - m_n^2}.
\ee
To proceed we decompose the sum over $m, n$ to three sums depending on
whether $m > n$, $m < n$ or $m = n$.
We carry out the sum for $m > n$ as follows
\bea
&&\sum_{n = 1}^{\infty} \sum_{m = n+1}^{\infty} c_m c_n \frac{m_m^2 \log m_m^2 - m_n^2 \log m_n^2}{m_m^2 - m_n^2} \nn \\
&=&\sum_{n=1}^{\infty} \sum_{r = 1}^{\infty} e^{z(2n+r)} \left( a(n+r) + \frac{a r}{e^{a r} - 1} \right) \nn \\
&=&a^2 \sum_{n=1}^{\infty} e^{2nz} \left(\frac{n}{e^{-z} - 1} + \frac{e^{-z}}{(e^{-z} - 1)^2} \right) +
a^2 \sum_{n=1}^{\infty}\sum_{r=1}^{\infty}\sum_{s=1}^{\infty} re^{z(2n+r)} e^{-a rs} \nn \\
&=&a^2 \left( \frac{e^{-2z}}{(e^{-z} - 1)(e^{-2z} - 1)^2} + \frac{e^{-z}}{(e^{-z} - 1)^2(e^{-2z} - 1)} +
\frac{1}{e^{-2z} - 1} \sum_{r=0}^{\infty} \frac{1}{4 \sinh^2((a(r+1) - z)/2)}  \right). \nn \\
\eea
From the 2nd line to the 3rd line,
we used the expansion
\be
\frac{1}{e^{a r} - 1} = \sum_{s=1}^{\infty} e^{-a rs}.
\ee
Note that all infinite series appearing in the derivation
are convergent for $z < 0$.
The last line is an expression which can be analytically continuated to $z_0 > 0$,
as long as (\ref{z0-cond}) is satisfied.
Obviously, the sum over $m < n$ gives the same value as the sum over $m > n$.
The sum over $m = n$ is straightforward
\be
\sum_{n = 1}^{\infty} c_n^2 \log m_n^2 = \sum_{n = 1}^{\infty} a n e^{2zn}
= \frac{a}{4 \sinh^2(z)}.
\ee
Again the last expression is well defined for $z \rightarrow z_0 > 0$.

\subsubsection{Generic diagrams}

From the explicit expressions (\ref{k20}) and (\ref{k2m}) for one-loop amplitudes,
we see that calculations in a similar fashion should
lead to a similar conclusion.
Though (\ref{k20}) and the second term in (\ref{k2m}) is more complicated than
the first term of (\ref{k2m}),
the difference is merely technical.
It suffices for our purpose to consider the first term in (\ref{k2m})
restricted to $0 < n_1 < n_2 < \cdots < n_V$ for an arbitrary choice of $(r_1, r_2, \cdots, r_V)$
\bea
&&\sum_{n_1=1}^{\infty}\sum_{n_2=n_1+1}^{\infty}\cdots\sum_{n_V=n_{V-1}+1}^{\infty}
c_{n_1}c_{n_2}\cdots c_{n_V} m_{n_1}^{-2r_1}m_{n_2}^{-2r_2}\cdots m_{n_V}^{-2r_V} \nn \\
&=&\sum_{s_1=1}^{\infty}\sum_{s_2=1}^{\infty}\cdots\sum_{s_V=1}^{\infty}
e^{z(Vs_1 + (V-1)s_2 + \cdots s_V)} e^{-a(r'_1 s_1 + r'_2 s_2 + \cdots r'_V s_V)} \nn \\
&=&\frac{1}{e^{r'_1 a-Vz} - 1}\frac{1}{e^{r'_2 a-(V-1)z} - 1}\cdots\frac{1}{e^{r'_V a-z} - 1},
\eea
where $r'_i = \sum_{j=i}^V r_j$.
Therefore we need
\be \label{irrational}
m z_0 - n a \neq 0 \quad \mbox{for} \quad m, n \in \mathbb{Z},
\ee
which includes (\ref{z0-cond}) as a special case,
for analytic continuation to work.
The condition (\ref{irrational}) is equivalent to say that
the ratio of $z_0$ to $a$ is irrational.
Through computations of many Feynman diagrams of higher loops,
we believe that this is also the sufficient condition for convergence.
A more rigorous proof will be presented in another paper.

While $z_0$ and $a$ can be chosen
so that their ratio is irrational,
the ratio can be appoximated with arbitrary accuracy by rational numbers.
If $n/m$ ($m, n$ are integers) is a good approximation of $z_0/a$,
the diagrams which diverge at $mz_0 = na$ will be large numbers.
Since better approximation of $z_0/a$ are in general reached
when $m$ and $n$ are larger integers,
and Feynman diagrams with larger ($m$, $n$) in their diverging condition $mz_0 = na$
involve more internal lines and hence more interaction vertices,
they are also suppressed by larger powers of the coupling constant $\lam$.
For $\lam$ sufficiently small,
the infinite-vertex limit of Feynman diagrams might be finite,
but generically we expect the perturbative expansion
of our theory to have a less convergent behavior than an ordinary theory.
On the other hand, even an ordinary field theory does not have
a convergent perturbative expansion,
which is actually an asymptotic expansion.
The situation is not qualitatively different.

\section{Discussion}

\subsection{More about the propagator}

The propagator (\ref{f}) is a high energy correction to
an ordinary $\phi^4$ theory with the propagator
\be
c_0/(k^2 - m_0^2)
\ee
if $m_1^2 \gg m_0^2$.
This is true if $z_0$ is close to $0$ and $a$ is not.
For example, if $z_0 = (2\pi^2)^{-1}$ and $a = 2$
(their ratio is irrational),
then $m_1^2$ is about 130 times bigger than $m_0^2$.
We should also rescale the propagator by field redefinition
so that it becomes
\be
\frac{1}{k^2 - m_0^2} + \frac{1}{c_0}\sum_{n=1}^{\infty}\frac{c_n}{k^2 - m_n^2}.
\ee

The propagator defined by (\ref{cnexp}), (\ref{mnexp}),
which we studied in some detail in this paper,
is not the only choice for our purposes.
In general, to construct a solution for (\ref{cm}) and (\ref{c}),
we first find an infinite series of positive numbers
which add up to a negative number or zero.
Then we re-adjust the values of $c_0$,
such that $\sum_n c_n = 0$.
Next we try to find $m_n$'s such that $\sum_n c_n m_n^2$
is again an infinite series of positive numbers adding up to a negative number or zero.
Then, by re-adjusting the first few terms of $m_n$,
we can satisfy $\sum_n c_n m_n^2 = 0$.

Here we give another interesting example:
\be \label{zeta}
c_n = n^{2z}, \quad m_n = n^2 \quad \mbox{for} \quad n = 1, 2, 3, \cdots.
\ee
Using zeta functions, we find
\be
\sum_{n=1}^{\infty} c_n m_n^{2r} = \zeta(-2(2r+z)).
\ee
This is zero for all $r \geq 1$ if $z$ is a positive integer.
Hence (\ref{cnmn}) is satisfied for {\em all} $r \geq 1$!
The propagator defined by (\ref{zeta}) might be able
to regularize UV-divergences of arbitrary orders.

An important point in constructing the series $\{(c_n, m_n)\}$
is that the prescription for analytic continuation should be
well defined without ambiguity.
If we choose $c_n = n^{2z}$ and $m_n^2 = e^{an}$ for example,
the sum $\sum_n c_n m_n^{2r}$ is well defined only for $r < 0$.
It is tempting to treat $a$ as a variable like $z$,
and take $a > 0$ for $r < 0$ and $a < 0$ for $r > 0$.
But then a Feynman diagram will need to be decomposed
into two parts with different values of $a$.
As this decomposition can be ambiguous,
the result will also be ambiguous.
As a rule one should not analytically continuate any parameter
in the definition of $m_n^2$,
since they appear in Feynman diagrams with both positive and negative powers.
On the other hand, since $c_n$ always comes in the product
$(c_{n_1}c_{n_2}\cdots c_{n_I})$,
the parameter $z$ in $c_n$ can always be analytically continuated
in the same way for all Feynman diagrams,
as long as $m_n$'s are chosen such that they do not diverge faster than $c_n$
when $n \rightarrow \infty$.

\subsection{Comments on unitarity}

Unitarity is almost immediate in our formulation.
The only potential problem is that the sum (\ref{imM}),
as a sum over physical probabilities,
must be positive.
We don't want the magic of turning infinitely many positive numbers
into a negative number to happen here.
Fortunately, energy conservation saves us from this catestrophy.
For a finite center of mass energy $E$,
only a finite number of poles can be excited if the masses $m_n$
approach to infinity when $n\rightarrow \infty$.
This holds for both examples (\ref{cnexp}), (\ref{mnexp}) and (\ref{zeta}).
The sum is always truncated at a finite $n$,
and all propagating modes at $m_n$ with $m_n > E$ can be ignored.
For our one-loop diagram Fig. 1(b), we can simplify (\ref{imM}) as
\be
\frac{1}{32\pi^2}\sum_{m,n}c_m c_n \int_0^{\infty} dk \frac{k}{p\sqrt{k^2 + m_m^2}}
\left(\Theta(E_m(k) + E_{n+}(k) - p_0) - \Theta(E_m(k) + E_{n-}(k) - p_0) \right),
\ee
where $k$ and $p$ are the norm of the spatial momenta denoted by $k$ and $p$ above,
and
\be
E_m(k) = \sqrt{k^2 + m_m^2}, \quad E_{n\pm}(k) = \sqrt{(k\pm p)^2 + m_n^2},
\ee
and $\Theta$ is the step function which is zero or one depending on whether
its argument is negative or positive.
Due to the step functions, the integrand is non-vanishing only if
\be
E_m + E_{n+} > p_0 > E_m + E_{n-},
\ee
which implies that
\be
p_0 > m_m, \quad \mbox{and} \quad p_0 > m_n.
\ee
Thus we can cut off the sum to a finite sum as we have argued,
and a finite sum of positive numbers is always positive.

\subsection{Analytic continuation}

It may appear unnatural that we are using some mathematical trick to replace
a sum of positive numbers by a negative number.
This may eventually turn out to be an artifect of the perturbation theory,
once we understand the theory better.
A well-known example in string theory
arises in the computation of Virasoro algebra as
\be
\sum_n n = \left. \sum_n n^{-s}\right|_{s = -1} = \zeta(-1) = - \frac{1}{12}.
\ee
Clearly analytic continuation of $s$ is used.
Our theory is analogous to the Feynman diagram computation of this quantity.
But in this case there is a more rigorous formulation which
does not reply on the analytic continuation.
At this moment we can only hope that a more rigorous formulation
will be found in the future.
For the time being,
our formulation is just a prescription to construct a UV-finite
quantum field theory which is unitary to all orders and all energies.
The analytic continuation is merely a regularization scheme which gets rid of
UV divergences for us.
In this sense, it is a priori possible that the final (finite) number corresponding
to a Feynman diagram depends on the regularization we use
for the loop momentum integral
(4-momentum cutoff, 3-momentum cutoff or dimensional regularization).
(Recall that we have to first regularize the integral over loop momentum
and then take the limit of $\Lam\rightarrow \infty$ or $\eps \rightarrow 0$
after summing over the infinite series.)
It is intriguing that for Fig. 1(a)
the final result is independent of the regularization we choose.
(See Sec. \ref{UVdiv}.)
It remains to be seen whether this is a generic feature for all Feynman diagrams.

\subsection{Generalizations}

It is straightforward to extend this framework to fermions.
For a fermionic field theory with UV divergences of order $D$,
we shall modify its propagator
\be
\frac{i}{\ks + m_0} \rightarrow \sum_n \frac{ic_n}{\ks + m_n},
\ee
where the parameters should satisfy
\be \label{fermion}
\sum_n c_n m_n^r = 0 \quad \mbox{for} \quad r = 0, 1, 2, \cdots, D,
\ee
and $c_n > 0$.

A closely related theory of fermions was considered by Itzhaki \cite{Itzhaki}.
Instead of modification of the propagator, his model consists of
infinitely many fields with standard propagators.
There are a number of $c_n$ of fields with the same mass $m_n$
for each positive integer $n$,
and the interaction among all fields is given in the Lagrangian density
as a polynomial of $(\sum_n \psi_n)$.
The same constraints (\ref{fermion}) are imposed to eliminate UV divergences.
His theory is essentially the fermionic analogue of (\ref{Ln}),
except that the analytic continuation is ambiguous
for his choice of parameters $c_n$, $m_n$.
(This ambiguity was mentioned above.)

For a long time string theory has been the only candidate
which admits a perturbation theory of gravity that is
both perturbatively UV-finite and unitary.
It will be very exciting to explore whether the ideas of this work
can be extended to describe gauge theories or even gravity.

\section*{Acknowledgement}

The authors thank Jiunn-Wei Chen, Chong-Sun Chu, Hiroyuki Hata,
Takeo Inami, Satoshi Iso, Yutaka Matsuo, Sergei Nedelko, Ryu Sasaki,
Richard Woodard and people in the string theory group in Taiwan.
This work is supported in part by
the National Science Council, Taiwan, R.O.C.

\end{document}